\documentclass[numberedheadings]{aipproc}
\layoutstyle{6x9}
\newcommand{\nc}{\newcommand}
\nc{\mb}[1]{\makebox[#1]{}}
\nc{\CC}{{\scriptscriptstyle CC}}
\nc{\NC}{{\scriptscriptstyle NC}}
\nc{\V}{{\rm v}}
\nc{\W}{{\scriptscriptstyle W}}
\nc{\X}{{\scriptscriptstyle X}}
\nc{\Z}{{\scriptscriptstyle Z}}
\nc{\CS}{{\scriptscriptstyle CS}}
\nc{\DY}{{\scriptscriptstyle DY}}
\nc{\PW}{{\scriptscriptstyle PW}}
\nc{\SB}{{\scriptscriptstyle SB}}
\nc{\CSV}{{\scriptscriptstyle CSV}}
\nc{\GLS}{{\scriptscriptstyle GLS}}
\nc{\CIB}{{\scriptscriptstyle CIB}}
\nc{\PT}{{\scriptscriptstyle PT}}
\nc{\IE}{{\it i.e.,\ }}
\nc{\EG}{{\it e.g.,\ }}
\nc{\EA}{{\it et al.\ }}
\nc{\AH}{{\it ad hoc\ }}
\nc{\CHPT}{{$\chi_{\PT}$\ }}

\newcommand{\inv}{\left. g^{(0)}_A \right|_{\rm inv}}

\newcommand{\Frac}[2]%
{{\textstyle \frac{\mbox{\footnotesize $#1$}\rule[-0.9mm]{0mm}{1mm}}%
{\mbox{\footnotesize $#2$}\rule{0mm}{3.1mm}}}}
\nc{\st}{\scriptstyle}
\nc{\sst}{\scriptscriptstyle}
\nc{\mco}{\multicolumn}
\nc{\epp}{\epsilon^{\prime}}
\nc{\vep}{\varepsilon}
\nc{\ra}{\rightarrow}
\nc{\ppg}{\pi^+\pi^-\gamma}
\nc{\nuN}{{\nu N_0}}
\nc{\nubN}{{\overline{\nu} N_0}}
\nc{\snuNC}{{\langle \sigma^{\nuN}_{\NC}\rangle }}
\nc{\snubNC}{{\langle \sigma^{\nubN}_{\NC}\rangle }}
\nc{\snuCC}{{\langle \sigma^{\nuN}_{\CC}\rangle }}
\nc{\snubCC}{{\langle \sigma^{\nubN}_{\CC}\rangle }}
\nc{\Rnu}{{R^{\nu}}}
\nc{\Rnub}{{R^{\overline{\nu}}}}
\nc{\sintW}{{\sin^2 \theta_{\W} }}
\nc{\MS}{{\overline{MS}}}
\nc{\vp}{{\bf p}}
\nc{\rz}{{1\over \rho_0^2}}
\nc{\ko}{K^0}
\nc{\kb}{\bar{K^0}}
\nc{\al}{\alpha}
\nc{\ab}{\bar{\alpha}}
\nc{\be}{\begin{equation}}
\nc{\ee}{\end{equation}}
\nc{\bea}{\begin{eqnarray}}
\nc{\eea}{\end{eqnarray}}

\begin{document}
\title{Neutrino Physics Without Oscillations}
\author{Anthony W. Thomas}{address={Special Research Centre for the 
Subatomic Structure of Matter, \\
University of Adelaide, Adelaide SA 5005, Australia}} 

\begin{abstract}
There is a tremendous potential for neutrinos to 
yield valuable new information about strongly interacting systems. Here 
we provide a taste of this potential, beginning with the existence of a 
rigorous sum rule for the proton and neutron spin structure functions 
based on the measurement of the flavor singlet axial charge of the nucleon.
We also comment on the NuTeV report 
of a 3$\sigma$ deviation of the value of $\sin^2 \theta_W$ measured in 
neutrino (and anti-neutrino) scattering from that expected within the 
Standard Model. 
\end{abstract}

\maketitle

\section{Introduction}
Apart from rather old bubble chamber experiments we have no data for neutrino
interactions with free protons. In spite of this, neutrino measurements on 
nuclei have given very important information on the internal structure of 
the nucleon. For example, this is the best source of information on the shape 
of the anti-quark parton distribution functions -- even though, as we mention 
below, the understanding of nuclear corrections for neutrino beams is in a 
very rudimentary state. The axial form factor of the nucleon provides important
guidance in dealing with low energy interactions of pions with nucleons and 
the nucleon-nucleon force. In comparison with the axial and electromagnetic 
transition form factors to baryon excited states it can also yield new insight 
into the dynamics of hadron structure \cite{Thomas:kw}. 

Sum rules play a crucial role in strong interaction physics as, ideally, they 
are model independent -- linking different experiments. The failure of such 
a sum rule indicates a serious change to our understanding of the strong 
interaction -- the failure of some symmetry or of QCD itself.
The famous Ellis-Jaffe sum rule, which failed in the so-called ``spin crisis'',
was not in this category as it required a dynamical assumption (the absence 
of strange quarks in the nucleon). On the other hand, it is not widely 
appreciated that developments in the application of Witten's renormalization 
group methods \cite{witten} to the flavor singlet 
axial axial charge of the nucleon mean 
that we {\it do} now have a rigorous sum rule for the leading twist spin 
structure functions of the nucleon \cite{Bass:2002mv,Bass:2002wy}. 
This is discussed in Sect. 2, where the 
importance of measuring neutrino-nucleon elastic scattering is emphasised.

In Sect. 3 we discuss recent developments in the interpretation of 
the reported 3$\sigma$ deviation from the Standard Model expectation for 
$\sin^2 \theta_W$ in neutrino nucleus scattering. There is a recent model 
independent result for the correction associated with charge symmetry
violation in the parton distribution functions which reduces the anomaly to 
no more than 2$\sigma$ \cite{Londergan:2003ij}. 
In addition, there are major issues concerning our 
understanding of shadowing in neutrino interactions with nuclei which urgently 
need attention -- both in connection with the NuTeV claim \cite{Miller:2002xh} 
and more generally 
in connection with the possible systematic errors in our knowledge of the 
flavor dependence of parton disributions.

\section{A Rigorous Sum Rule for Spin Dependent Deep Inelastic Scattering}
In polarised deep inelastic scattering at a scale, $Q^2$, where three flavors
of quark are ``active'', the first moment of
the nucleon spin structure function, $g_1$, may be written as
a linear combination of the iso-triplet, SU(3) 
octet and flavour singlet, scale-invariant axial charges of
the nucleon \cite{kod,mink,larin,alta}:
\begin{equation}
\int_0^1 dx \ g_1^p(x,Q^2) =
\Biggl( {1 \over 12} g_A^{(3)} + {1 \over 36} g_A^{(8)} \Biggr)
C_{NS}(Q^2)
+ {1 \over 9} g_A^{(0)}|_{\rm inv} C_S(Q^2)
+ {\cal O}({1 \over Q^2}) \, .
\label{eq:SR}
\end{equation}
Here $C_{NS}$ and $C_S$ are, respectively, the flavour non-singlet 
and flavour singlet Wilson coefficients, which have been evaluated to 
three-loops in Ref.~\cite{larin}:
\begin{eqnarray}
C_{NS}(Q^2) &=& 
\left[1 - \frac{\alpha_s}{\pi} 
        - 3.58 \left(\frac{\alpha_s}{\pi} \right)^2
        - 20.21 \left(\frac{\alpha_s}{\pi} \right)^3 \right] 
 \nonumber \\
C_S (Q^2)   &=&
\left[1 - \frac{\alpha_s}{3 \pi} 
        - 0.55 \left(\frac{\alpha_s}{\pi} \right)^2 
        - 4.45 \left(\frac{\alpha_s}{\pi} \right)^3 \right] \, . 
\label{Lar}
\end{eqnarray}
The isotriplet axial charge, $g_A^{(3)}$:
\begin{equation}
2 m s_{\mu} \ g_A^{(3)} =
\langle p,s |
\left(\bar{u}\gamma_\mu\gamma_5u - \bar{d}\gamma_\mu\gamma_5d \right)
| p,s \rangle \, ,
\end{equation}
is measured in neutron beta-decay, while the octet axial charge, $g_A^{(8)}$:
\begin{equation}
2 m s_{\mu} \ g_A^{(8)} =
\langle p,s |
\left(\bar{u}\gamma_\mu\gamma_5u + \bar{d}\gamma_\mu\gamma_5d
                   - 2 \bar{s}\gamma_\mu\gamma_5s\right)
| p,s \rangle \, ,
\end{equation}
is measured in hyperon beta-decays. 
The Ellis-Jaffe sum rule \cite{ej} was based on the 
dynamical assumption that the strange quark content of the nucleon was 
negligible, in which case the scale-invariant, flavor singlet axial charge 
would be equal to the octet axial charge. 
The failure of that sum rule \cite{smc,hermes,e143,e154} led to 
a completely new appreciation of the role of the axial anomaly in QCD, as well 
as the possibility that a substantial fraction of the spin of the proton might 
reside on polarized glue.

Recently, Bass, Crewther, Steffens and Thomas (BCST) \cite{Bass:2002mv} 
have shown 
how the 3-flavour, scale-invariant, flavour-singlet axial charge of the 
nucleon can be determined independently by systematically correcting  
elastic neutrino-proton scattering data for heavy-quark 
contributions. With this result we have a {\bf rigorous sum rule} 
relating deep inelastic scattering in the Bjorken region of high energy and 
momentum transfer to three independent, low energy measurements. As with the 
Bjorken sum rule, the verification of this new proton spin sum rule is a 
crucial test of QCD itself.

The scale-invariant flavour singlet axial charge, $g_A^{(0)}|_{\rm inv}$, 
is defined by \cite{mink,larin}:
\begin{equation}
2m s_\mu \inv =
\langle p,s|
{\cal S}_{\mu} (0) |p,s\rangle
\label{e}
\end{equation}
where
\begin{equation}
{\cal S}_\mu(x) = E(g)J^{GI}_{\mu5}(x)
\label{f}
\end{equation}
is the product of the gauge-invariantly renormalized singlet 
axial-vector operator
\begin{equation}
J^{GI}_{\mu5} = \left(\bar{u}\gamma_\mu\gamma_5u
                  + \bar{d}\gamma_\mu\gamma_5d
                  + \bar{s}\gamma_\mu\gamma_5s\right)_{GI}
\label{aa}
\end{equation}
in three flavour QCD
and the renormalisation group factor
\begin{equation}
E(g) = \exp \int^g_0 \! dg'\, \gamma(g')/\beta(g') \, .
\label{a}
\end{equation}
Here
$\beta(g)$ and $\gamma(g)$ are the
Callan-Symanzik functions associated with the
gluon coupling constant $g$ and the composite operator
$J^{GI}_{\mu5}$. Finally then, the renormalization group invariant, 
singlet axial charge is (for three active flavors):
\begin{eqnarray}
g_A^{(0)}\bigr|_{\rm inv} &=&  E_3(\alpha_3)
  \bigl(\Delta u + \Delta d + \Delta s\bigr)_3   \nonumber \\
&=& \bigl(\Delta u + \Delta d + \Delta s\bigr)_{\rm inv} \, .
\label{inv2}
\end{eqnarray}

The non-perturbative factor $E(g)$ in Eq.(\ref{f}) 
arises naturally from the coefficient
function of $J^{GI}_{\mu5}$ in the product of electromagnetic currents
$J_\alpha(x)J_\beta(0)$ at short distances, $x_\mu \sim 0$. It
compensates for the scale dependence of $J^{GI}_{\mu5}$ caused by the
anomaly \cite{adler,bell,koeb,rjc}
in its divergence
\begin{equation}
\partial^\mu J^{GI}_{\mu5}
= 2f\partial^\mu K_\mu + \sum_{i=1}^{f} 2im_i \bar{q}_i\gamma_5 q_i \, ,
\end{equation}
where $K_\mu$ is a renormalized version of the gluonic Chern-Simons 
current, and the number of flavours, $f$, is three. 
A consequence of the renormalization-scale invariance of 
${\cal S}_\mu$ is that its spatial components have operator charges
\begin{equation}
S_i(t) = \int\! d^3x\, {\cal S}^i(t,{\bf x}) \, ,
\end{equation}
which satisfy an equal-time algebra \cite{mink}
\begin{equation}
[S_i(t),S_j(t)] = i\varepsilon_{ijk}S_k(t) \, ,
\end{equation}
characteristic of spin operators, whereas the operator charges
$\int d^3x J_{i5}^{GI}(t, {\bf x})$ do not \cite{bt}.

Elastic $\nu p$ scattering measures the neutral-current 
axial charge $g_A^{(Z)}$
\begin{equation}
2 m s_{\mu} \ g_A^{(Z)} =
 \langle p,s|J^Z_{\mu5}|p,s\rangle \, ,
\end{equation}
where
\begin{equation}
J_{\mu5}^Z\ 
=\ \Frac{1}{2} \biggl\{\,\sum_{q=u,c,t} - \sum_{q=d,s,b}\,\biggr\}\:
        \bar{q}\gamma_\mu\gamma_5q \, .
\end{equation}
As this is a six flavour quantity one must correct for
the heavy flavours $t,b$ and $c$ which do not contribute to deep
inelastic scattering for $Q^2$ below (say) 10 GeV$^2$.
By applying the renormalization group techniques of Witten, and in particular 
by introducing appropriate ``matching functions'', BCST were able to sum all 
large logarithms appearing to NLO (the technique can be applied
systematically to any order) in the heavy quark masses, 
with the result \cite{Bass:2002mv}:
\begin{eqnarray}
2 g_A^{(Z)} &=& \bigl(\Delta u - \Delta d - \Delta s\bigr)_{\rm inv}
 +\hspace{0.2cm} {\cal P}\hspace{0.1cm}\bigl(
\Delta u + \Delta d + \Delta s\bigr)_{\rm inv}
\nonumber \\
&+& O(m_{t,b,c}^{-1}) \, ,
\label{g2}
\end{eqnarray}
where ${\cal P}$ is a polynomial in the
running couplings
\begin{eqnarray}
{\cal P} &=& \frac{6}{23\pi} \bigl(\alpha_b-\alpha_t\bigr)
             \Bigl\{1 + \frac{125663}{82800\pi} \alpha_b
                      + \frac{6167}{3312\pi} \alpha_t
                      - \frac{22}{75\pi} \alpha_c\Bigr\}
\nonumber \\
&-& \frac{6}{27\pi} \alpha_c
                      - \frac{181}{648 \pi^2}\alpha_c^2
                      + O\bigl(\alpha_{t,b,c}^3\bigr) \, .
\label{eq:PP}
\end{eqnarray}
Here $\alpha_t$, $\alpha_b$ and $\alpha_c$ are rigorously defined 
simultaneous running couplings. The factor $\alpha_b-\alpha_t$ 
ensures that all contributions from $b$ and $t$ quarks cancel 
{}for $m_t=m_b$. 

The final step in deriving the sum rule is then to realize that 
\be
(\Delta u - \Delta d - \Delta s\bigr)_{\rm inv} = 
g_A^{(3)} + \frac{1}{3} g_A^{(8)} - \frac{1}{3}
g_A^{(0)}|_{\rm inv} \, .
\label{eq:G0}
\ee
Using Eqs.(\ref{g2}), (\ref{eq:PP}) and (\ref{eq:G0}), one can extract
the value of $g_A^{(0)}|_{\rm inv}$ and Eq.(\ref{eq:SR}) is then a 
rigorous sum rule. As a result, the accurate measurement of the $Z^0$
axial coupling to the proton (or neutron) should be an extremely high
priority.

\section{The NuTeV Anomaly}
In 1973, Paschos and Wolfenstein \cite{Paschos:1972kj} derived an expression 
relating the ratio of neutral-current and charge-changing neutrino 
interactions on isoscalar targets to the Weinberg angle:
\be
 R^- \equiv { \rz \left( \snuNC - \snubNC \right) \over 
 \snuCC - \snubCC } = {1\over 2} - \sintW  .
\label{eq:PasW} 
\ee  
In Eq.(\ref{eq:PasW}), $\snuNC$ and $\snuCC$ are respectively the 
neutral-current and charged-current inclusive, total 
cross sections for neutrinos on an 
isoscalar target.  The quantity $\rho_0 \equiv M_{\W}/(M_{\Z}\,\cos 
\theta_{\W})$ is one in the Standard Model.  
The NuTeV group recently measured neutrino charged-current and 
neutral-current cross sections on iron \cite{Zeller:2001hh}, finding 
$\sintW = 0.2277 \pm 0.0013~({\rm stat}) \pm 0.0009 ~({\rm syst})$.  This 
value is three standard deviations above the measured fit to 
other electroweak processes, $\sintW = 0.2227 \pm 0.00037$ 
\cite{Abbaneo:2001ix}.  
The discrepancy between the NuTeV measurement and determination of the 
Weinberg angle from electromagnetic measurements is surprisingly large, 
and may constitute evidence of physics beyond the 
Standard Model.   

As the NuTeV experiment did not strictly measure the Paschos-Wolfenstein
ratio, there are a number of additional
corrections that need to be considered, such as shadowing
\cite{Miller:2002xh}, asymmetries in $s$ and $\bar{s}$ distributions
\cite{Cao:2003ny}, asymmetries in $c$ and $\bar{c}$ 
distributions \cite{Melnitchouk:1997ig}, charge asymmetric valence parton
distributions \cite{Londergan:2003ij,Londergan:2003pq}, and so on.
Reference \cite{Davidson:2001ji} provides an excellent summary of 
possible corrections to the NuTeV 
result from within and outside the Standard Model.  

\subsection{Model Independence of the Charge Symmetry Correction}
Londergan and Thomas recently calculated corrections to the NuTeV experiment 
arising from charge symmetry violation CSV) caused by the small 
difference of $u$ and $d$ quark 
masses \cite{Londergan:2003pq}. 
This calculation followed earlier work on CSV in parton 
distributions \cite{Sather:1991je,Rodionov:cg}, and 
involved calculating CSV distributions 
at a low momentum scale, appropriate to a valence-dominated quark model,
and using QCD evolution to generate the CSV
distributions at the $Q^2$ 
values appropriate for the NuTeV experiment. 
The result was a correction to the 
NuTeV result $\Delta R_{\CSV} \sim -0.0015$. This would reduce the
reported effect from 3 to 2 standard deviations. Following this, NuTeV  
reported their own estimate of the CSV parton distributions, using a rather 
different procedure \cite{NuTeV2}. They obtained  
a much smaller correction, $\Delta R_{\CSV} \sim +0.0001$. 
The large discrepancy between these two results suggested that the 
CSV correction might be strongly model dependent.

This question was recently resolved by Londergan and Thomas 
\cite{Londergan:2003ij}, as we now summarise.
The charge symmetry violating contribution to the Paschos-Wolftenstein ratio 
has the form 
\be 
 \Delta R_{\CSV} = \left[ 3\Delta_u^2 + \Delta_d^2 + {4\alpha_s \over 9\pi}
 \left(\bar{g}_L^2 - \bar{g}_R^2 \right) \right] \,\left[ {\delta U_{\V} - 
 \delta D_{\V}  \over 2(U_{\V} + D_{\V}) }\right] 
\label{eq:CSV} 
\ee
where 
\bea 
 \delta Q_{\V} &=& \int_0^1 \, x\,\delta q_{\V}(x) \,dx \nonumber \\ 
 \delta d_{\V}(x) &=& d_{\V}^p(x)- u_{\V}^n(x)~; \hspace{1.0cm}
 \delta u_{\V}(x) = u_{\V}^p(x)- d_{\V}^n(x) ~.
 \label{eq:CSVadd} 
\eea
The denominator in the final term in Eq.\ (\ref{eq:CSV}) gives the total 
momentum carried by up and down valence quarks, while the numerator 
gives the charge symmetry violating momentum 
difference -- for example, $\delta U_{\V}$, is the total momentum carried by up 
quarks in the proton minus the momentum of down quarks in the neutron.  
This ratio is completely independent of $Q^2$ and can be evaluated at
any convenient value.
  
Using an analytic approximation to the charge symmetry violating valence
parton distributions that was initially proposed
by Sather \cite{Sather:1991je}, 
one can evaluate Eq.(\ref{eq:CSV}) at a low scale,
$Q_0^2$, appropriate for a (valence dominated) quark or bag
model~\cite{Signal:yc,Schreiber:tc}. 
The advance over earlier work was to realize
that for NuTeV we need only the first moments of the CSV distribution
functions and these could be obtained analytically. The result for
the moment of the CSV down valence distribution, $\delta D_{\V}$, is
\bea 
 \delta D_{\V} &=& \int_0^1 \, x  \left[ -\frac{\delta M}{M} \frac{d}{dx} 
 (x d_{\V}(x)) - \frac{\delta m}{M} \frac{d}{dx} d_{\V}(x) \right] 
 \, dx \nonumber \\ &=& \frac{\delta M}{M} \int_0^1 \, x \, d_{\V}(x)\, dx + 
 \frac{\delta m}{M} \int_0^1 \, d_{\V}(x)\, dx = \frac{\delta M}{M} D_{\V} 
 + \frac{\delta m}{M} \, , 
\label{eq:intDv}
\eea   
while for the up quark CSV distribution it is 
\bea 
\delta U_{\V} &=& \frac{\delta M}{M} \left[ \int_0^1 \, x\, \left( - 
\frac{d}{dx}\left[ x u_{\V}(x)\right] + \frac{d}{dx} u_{\V}(x) \right) 
\, dx \right] \nonumber \\
&=& \frac{\delta M}{M} \left( \int_0^1 \,x \, u_{\V}(x)\, dx - \int_0^1 \,  
u_{\V}(x)\, dx \right) = \frac{\delta M}{M} \left( U_{\V} - 2 \right)
\, .
\label{eq:intUv}
\eea  
(Here $\delta M = 1.3$ MeV is the neutron-proton mass difference, and
$\delta m = m_d - m_u \sim 4$ MeV is the down-up quark mass difference.)

Equations (\ref{eq:intDv}) and (\ref{eq:intUv}) show that 
the CSV correction to the Paschos-Wolfenstein ratio depend only on 
the fraction of the nucleon momentum carried by up and down valence quarks.  
At no point do we have to calculate specific CSV distributions.  
At the bag model scale, $Q_0^2 \approx 0.5$ GeV$^2$, the momentum fraction 
carried by down valence quarks,   
$D_{\V}$, is between $0.2-0.33$, and the total momentum fraction carried 
by valence quarks is $U_{\V} + D_{\V} \sim .80$.  From Eqs. (\ref{eq:intDv}) 
and (\ref{eq:intUv}) this gives $\delta D_{\V} \approx 0.0046$, 
$\delta U_{\V} \approx -0.0020$.  
Consequently, evaluated at the quark model scale, the CSV correction to the 
Paschos-Wolfenstein ratio is 
\be 
 \Delta R_{\CSV} \approx 0.5 \left[ 3\Delta_u^2 + \Delta_d^2 \right] 
 \,{\delta U_{\V} - \delta D_{\V} \over 2(U_{\V} + D_{\V}) } \approx 
 -0.0020 \, .
\label{eq:CSVnut}
\ee
Once the CSV correction has been calculated at some quark model scale,
$Q^2_0$, the ratio appearing in Eq.~(\ref{eq:CSV}) is independent of
$Q^2$, because both the numerator and
denominator involve the same moment of a non-singlet distribution. 
(Note that we have not included the small QCD radiative 
correction in Eq.(\ref{eq:CSVnut}).) 

We stress that both Eqs.~(\ref{eq:intDv}) and (\ref{eq:intUv}) are only
weakly dependent on the choice of quark model scale -- through the
momentum fractions $D_{\V}$ and $U_{\V}$, which are slowly varying functions
of $Q^2_0$ and, in any case, not the dominant terms in
those equations. This, together with the $Q^2$-independence of the
Paschos-Wolfenstein ratio (Eq.~(\ref{eq:CSV})) under QCD
evolution, explains why the previous results, obtained by Londergan and
Thomas with different models and at different 
values of $Q^2$~\cite{Londergan:2003pq},
were so similar. Finally, Londergan and Thomas also demonstrated that
the acceptance function calulated by NuTeV does not introduce any
significant model dependence to this result.

\subsection{Higher-Twist Shadowing Corrections}
Although the average $Q^2$ of the NuTeV measurement is quoted as 16
GeV$^2$, at small $x$ the typical $Q^2$ is much lower and one needs to
beware of possible higher-twist effects. In particular, studies of the
muon nucleus scattering in a similar momentum transfer region suggest
that vector meson dominance (VMD) processes can produce substantial
nuclear corrections \cite{Kwiecinski:ys,Melnitchouk:1995am,Melnitchouk:2002ud}. 
It is especially important that
there are extensively studied differences in shadowing between photon
and charged-current neutrino interactions with nuclei 
\cite{Boros:1998es,Boros:1998mt}. 
These were important in reducing the apparently large charge synmmetry
violation in an earlier NuTeV measurement \cite{Boros:1999fy}. 

As noted by Miller and Thomas \cite{Miller:2002xh}, 
the same reasoning leads one 
to conclude that there should be a substantial difference between
the VMD shadowing corrections for neutral and charged current neutrino
scattering. Such a difference was not considered 
in the analysis of the NuTeV data.
A priori, this correction is at least as large as the reported anomaly.
It is difficult to estimate the systematic error associated with this as
the NuTeV analysis requires that one model separately the ratios of neutral to
charged current cross sections for neutrinos and anti-neutrinos and the
input parton distributions are derived without higher-twist shadowing
corrections from a variety of sources including electron, muon and
neutrino data on protons, deuterons and nuclei. This needs a great deal
more work before one can demonstrate that the problem is under control
at the required level of accuracy.

\section{Summary}

We have explained the developments in systematically correcting for heavy quark
contributions to the flavor singlet axial charge of the nucleon measured
in neutrino, neutral current scattering from the nucleon, which mean
that we now have a rigorous spin sum rule. As a result it is now
imperative to find ways to measure $g_A^{(Z)}$ accurately.

In connection with the NuTeV anomaly we now have a robust prediction for 
the CSV contribution to the Paschos-Wolfenstein ratio.
It was possible to express the correction in terms  
of integrals which could be evaluated without ever specifying
the shapes of the CSV distributions. 
Despite the fact that parton charge symmetry
violation has not been directly measured experimentally, and that
parton CSV effects are predicted to be quite small, we have strong
theoretical arguments regarding both the sign and magnitude of these
corrections.  The CSV effects should make a significant contribution to
the value for the Weinberg angle extracted from the NuTeV neutrino
measurements, reducing the anomaly by at least one standard deviation. 
{}Finally, again in connection with the NuTeV anomaly, we noted the 
importance of understanding higher-twist shadowing corrections
associated with VMD and particularly the corresponding systematic errors
in our knowledge of parton distribution functions.

It is a pleasure to acknowledge important contributions to the topics
discussed here from S.~D.~Bass, R.~J.~Crewther, J.~T.~Londergan, G.~A.~Miller
and F.~Steffens. This work was supported by the Australian Research
Council and the University of Adelaide.

\end{document}